\documentclass[aps,prx,twocolumn,showpacs,superscriptaddress]{revtex4-1}

\usepackage{latexsym}
\usepackage{ifpdf}
\usepackage{graphicx}
\usepackage{epsfig}
\usepackage{amsmath, amssymb, bbm, mathrsfs}
\usepackage{multirow}
\usepackage{ulem}
\usepackage{fancyhdr}
\usepackage{color}
\usepackage{bm}
\usepackage{eurosym}
\usepackage{subfigure}
\usepackage{comment}
\usepackage{scrhack}
\usepackage{physics}
\usepackage{balance}
\usepackage{float}
\usepackage{flushend}
\usepackage{setspace}   % set line spacing

\usepackage[unicode=true,pdfusetitle,
 bookmarks=false,
 breaklinks=true,pdfborder={0 0 0},pdfborderstyle={},backref=false,colorlinks=true]
 {hyperref}
\hypersetup{
 colorlinks,linkcolor=red,citecolor=blue}

\newcommand{\beginsup}{%
        \setcounter{table}{0}
        \renewcommand{\thetable}{S\arabic{table}}%
        \setcounter{figure}{0}
        \renewcommand{\thefigure}{S\arabic{figure}}%
        \setcounter{equation}{0}
        \renewcommand{\theequation}{S.\arabic{equation}}%
        \setcounter{enumiv}{0}
        \renewcommand{\theenumiv}{S\arabic{enumiv}}%
        \setcounter{section}{0}
        \renewcommand{\thesection}{\Alph{section}}%
        \setcounter{subsection}{0}
        \renewcommand{\thesubsection}{\thesection.\arabic{subsection}}%
}

\begin{document}

\title{Cavity piezo-mechanics for superconducting-nanophotonic quantum interface}

\author{Xu Han}
\affiliation{Department of Electrical Engineering, Yale University, New Haven, Connecticut 06520, USA}

\author{Wei Fu}
\affiliation{Department of Electrical Engineering, Yale University, New Haven, Connecticut 06520, USA}

\author{Changchun Zhong}
\affiliation{Department of Applied Physics, Yale University, New Haven, Connecticut 06520, USA}
\affiliation{Yale Quantum Institute, Yale University, New Haven, Connecticut 06520, USA}

\author{Chang-Ling Zou}
\affiliation{Department of Electrical Engineering, Yale University, New Haven, Connecticut 06520, USA}

\author{Yuntao Xu}
\affiliation{Department of Electrical Engineering, Yale University, New Haven, Connecticut 06520, USA}

\author{Ayed Al Sayem}
\affiliation{Department of Electrical Engineering, Yale University, New Haven, Connecticut 06520, USA}

\author{Mingrui Xu}
\affiliation{Department of Electrical Engineering, Yale University, New Haven, Connecticut 06520, USA}

\author{Sihao Wang}
\affiliation{Department of Electrical Engineering, Yale University, New Haven, Connecticut 06520, USA}

\author{Risheng Cheng}
\affiliation{Department of Electrical Engineering, Yale University, New Haven, Connecticut 06520, USA}

\author{Liang Jiang}
\affiliation{Department of Applied Physics, Yale University, New Haven, Connecticut 06520, USA}
\affiliation{Yale Quantum Institute, Yale University, New Haven, Connecticut 06520, USA}

\author{Hong X. Tang}
\email{hong.tang@yale.edu}
\affiliation{Department of Electrical Engineering, Yale University, New Haven, Connecticut 06520, USA}
\affiliation{Yale Quantum Institute, Yale University, New Haven, Connecticut 06520, USA}

%\date{\today}

\begin{abstract}
Hybrid quantum systems are essential for the realization of distributed quantum networks. In particular, piezo-mechanics operating at typical superconducting qubit frequencies features low thermal excitations, and offers an appealing platform to bridge superconducting quantum processors and optical telecommunication channels. However, integrating superconducting and optomechanical elements at cryogenic temperatures with sufficiently strong interactions remains a tremendous challenge. 
Here, we report an integrated superconducting cavity piezo-optomechanical platform where 10-GHz phonons are resonantly coupled with photons in a superconducting and a nanophotonic cavities at the same time. 
Benefited from the achieved large piezo-mechanical cooperativity ($C_\mathrm{em}\sim7$) and the enhanced optomechanical coupling boosted by a pulsed optical pump, we demonstrate coherent interactions at cryogenic temperatures via the observation of efficient microwave-optical photon conversion.
This hybrid interface makes a substantial step towards quantum communication at large scale, as well as novel explorations in microwave-optical photon entanglement and quantum sensing mediated by gigahertz phonons.
\end{abstract}

\maketitle

\section{Introduction}

Combining the most advanced technologies in different regimes, hybrid quantum architectures are pivotal for the development of quantum information science \cite{Cirac1997,Kimble2008}. Recent few years have witnessed impressive progresses in two very important fields---quantum computing based on the state-of-the-art superconducting qubit technology \cite{Clarke2008,DevoretSchoelkopf2013,Kurpiers2018,Chou2018,ZhongCleland2019,Arute2019} and quantum telecommunication through low-loss optical photons \cite{Valivarthi2016,Yin2017,Ren2017,Boaron}. However, towards a future scalable quantum network, it is becoming increasingly urgent to interface the superconducting and the photonic modalities in a quantum coherent manner \cite{Lambert2019,Lauk2019}. To fully exploit the quantum advantages, such coherent superconducting-photonic interface must satisfy very stringent requirements: the signal inter-conversion must have not only a high efficiency but also low added noise.

Various physical platforms have been investigated to address these challenges in all aspects, including trapped irons or atoms \cite{Gard2017,Han_Atom}, magnonics \cite{Hisatomi}, electro-optics \cite{Tsang2010,Galy2016EO,Soltani2017EO,Fan2018}, optomechanics \cite{Tian2010OM,Barzanjeh2011OM,Wang2012OM,Bochmann2013,Bagci2014,Pitanti2015Painter,Zou_2016,Rueda2016,Vainsencher2016,Midolo2018OM,Andrews2014,Higginbotham2018}, etc. Among all the schemes, the optomechanical system is very appealing because strong photon-phonon coupling can be achieved in both microwave and optical domains to mediate efficient photon conversion. Benchmark demonstration of high-efficiency conversion has been recently achieved using a megahertz cavity electro-optomechanical system \cite{Andrews2014,Higginbotham2018}. But the low-frequency mechanical membrane inevitably suffers from large thermal noise even at millikelvin temperatures, and the 3D optical cavity design poses a limit to the scalability. 

On the other hand, gigahertz piezo-optomechanics \cite{Chan2011,Han2014,Han2015,Cohen2015,Vainsencher2016,Hong2017,Riedinger2018,Ramp2019} is advantageous in that thermal excitations ($\bar{n}_\mathrm{th}$) at high frequencies are significantly suppressed, $\bar{n}_\mathrm{th}\approx\frac{k_\mathrm{B}T}{\hbar\omega}$, where $k_\mathrm{B}$ is the Boltzmann constant, $\hbar$ is the reduced Planck constant, $\omega$ and $T$ are the frequency and the temperature. Moreover, these high-performance piezo-optomechanical (POM) micro-devices can be fully fabricated on a chip, offering great potential for integration and up-scaling.

\begin{figure*}
\begin{centering}
\includegraphics{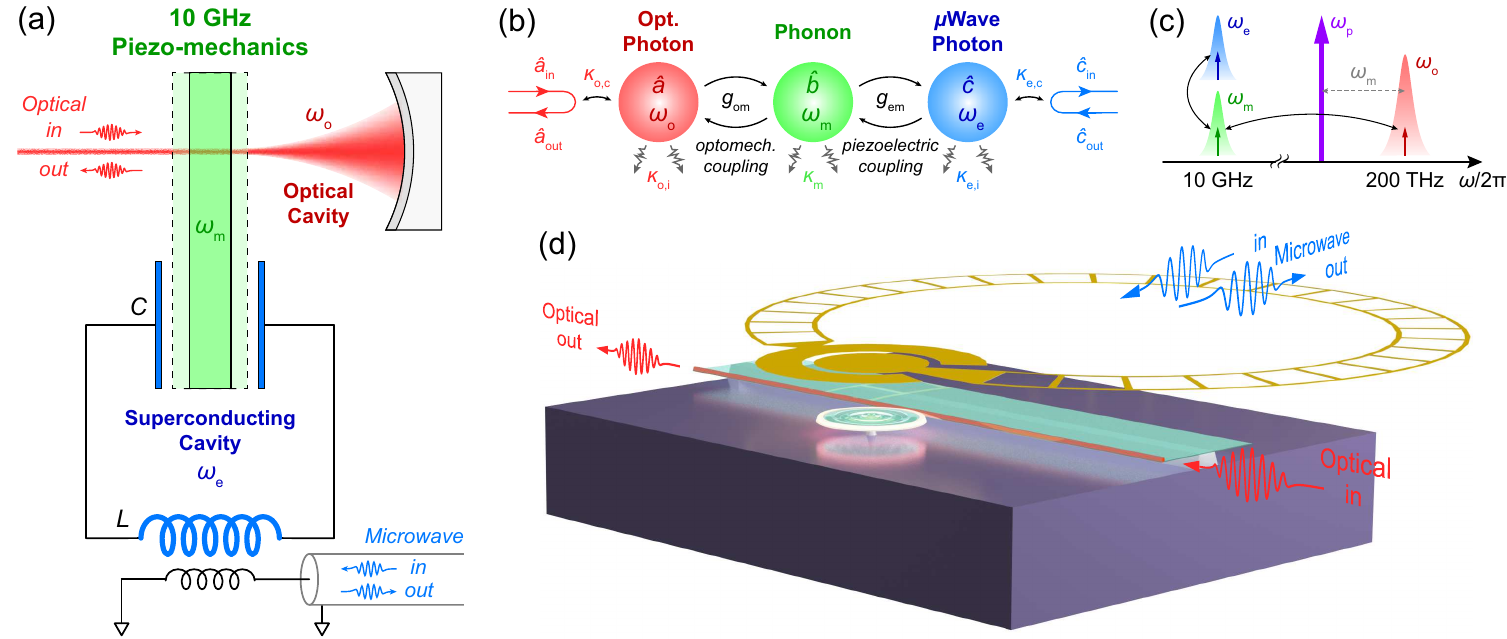}
\par\end{centering}
\caption{(a) Superconducting cavity piezo-optomechanical system. A 10-GHz mechanical resonator is simultaneously coupled with an optical and a superconducting cavities to achieve resonantly enhanced optomechanical and electromechanical interactions at the same time. The superconducting cavity is inductively coupled to a transmission line for microwave signal input and output. (b) Interaction diagram of the triple-resonance system. $g_\mathrm{em}$ and $g_\mathrm{om}$ denote the electromechanical and the cavity-enhanced optomechanical the coupling rates, respectively. (c) Photon conversion mechanism in the frequency domain. The blue, green, and red Lorentzian shapes stand for the microwave, mechanical, and optical resonances, respectively. An red-detuned optical pump is indicated as the purple arrow. (d) Experimental realization of the triply resonant superconducting piezo-optomechancial interface (schematic not to scale). A frequency-tunable superconducting ``Ouroboros" microwave resonator (yellow) is aligned and coupled with a piezo-optomechanical micro-disk through the piezoelectric effect. Microwave and optical photons are interconverted via cavity-enhanced interactions with 10-GHz phonons supported by the thickness mode of the micro-disk.}
\label{fig1}
\end{figure*}

However, it remains a tremendous challenge to efficiently interface an on-chip POM system with superconducting circuits. 
During past few years, great efforts have been made towards this goal. 
Bidirectional coherent conversion has been observed in room-temperature experiments by using a gigahertz POM crystal \cite{Vainsencher2016,Jiang2019}. Recently, such a POM crystal has been cooled to its ground state for low-noise microwave-to-optical conversion \cite{Forsch2019}.
Nevertheless, due to the lack of low-dissipation cavity-enhanced microwave coupling, in all existing realizations of POM converters, the achieved photon conversion efficiency is very limited ($\sim$\,$10^{-5}$ at room temperature \cite{Jiang2019} and $\sim$\,$10^{-10}$ at cryogenic temperature \cite{Forsch2019}).
Although superconducting cavity electromechanics has accomplished great successes \cite{Palomaki2013a,Han_2016,Gely2019}, combining superconducting and photonic devices is extremely difficult since superconductor absorbs light and light breaks superconductivity. 
It becomes even more challenging to integrate a superconducting cavity in the vicinity of a suspended optomechanical resonator with large mode overlap and minimized mode volume for enhanced interaction. Only a few potential designs have been proposed theoretically \cite{Zou_2016,Wu2019}, but no experimental realization has yet been achieved.

In this work, we demonstrate a triply resonant superconducting cavity POM interface that enables efficient microwave-optical (M-O) photon conversion. 
To our knowledge, this is the first realization of a superconducting POM converter that permits simultaneous cavity enhancement of photon-phonon interactions in both microwave and optical domains. 
By integrating a frequency-tunable superconducting cavity with a 10-GHz POM micro-disk, we are able to substantially enhance the microwave photon-phonon interaction and achieve a large electromechanical cooperativity ($C_\mathrm{em}\sim7$)---about three orders of magnitude improvement compared to previous POM converter demonstrations \cite{Jiang2019}. Combined with the implementation of a pulsed optical pump scheme to simultaneously boost the optomechanical interaction, we demonstrate coherent photon and phonon interactions via the observation of efficient bidirectional M-O photon conversion. 
The demonstrated superconducting-nanophotonic interface would not only advance the development of scalable quantum information networks, but also incorporate the state-of-the-art superconducting quantum technologies in the optical domain for breakthroughs in hybrid quantum systems, such as entangled M-O photon pair generation \cite{Zhong2019}, quantum repeaters \cite{Li2017}, and quantum metrology and sensing \cite{WangSun2019}.

\section{Hybrid cavity piezo-mechanics}

The concept of our triply resonant superconducting POM system is illustrated in Fig.\,\ref{fig1}(a). A high-frequency piezo-mechanical resonator at 10\,GHz is coupled to an optical cavity through radiation pressure force. At the same time, the mechanical mode is immerged in the electric field from the capacitor of an superconducting LC resonator. As a result, cavity-enhanced piezoelectric interaction between the mechanical and the microwave modes can be achieved. The interaction diagram of the system is shown in Fig.\,\ref{fig1}(b), where $\hat{a},\hat{b},\hat{c}$ are the annihilation operators of the optical, the mechanical, and the microwave modes, respectively, with resonant frequencies $\omega_\mathrm{o},\omega_\mathrm{m}$, and $\omega_\mathrm{e}$. The intrinsic dissipation rates of the three modes are denoted by $\kappa_\mathrm{o,i},\kappa_\mathrm{m}$, and $\kappa_\mathrm{e,i}$. Input and output photons are coupled to the optical and the microwave modes with respective external coupling rates $\kappa_\mathrm{o,c}$ and $\kappa_\mathrm{e,c}$, contributing to the total dissipation rates $\kappa_\mathrm{o}\equiv\kappa_\mathrm{o,i}+\kappa_\mathrm{o,c}$ and $\kappa_\mathrm{e}\equiv\kappa_\mathrm{e,i}+\kappa_\mathrm{e,c}$ of the two modes.

\begin{figure*}
\begin{centering}
\includegraphics{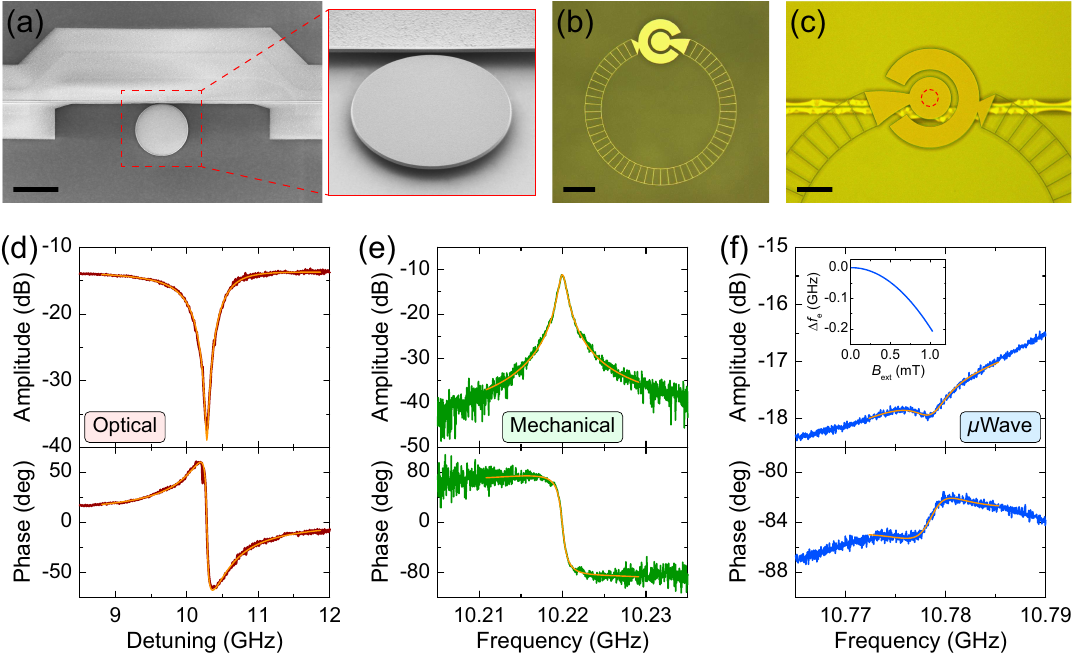}
\par\end{centering}
\caption{(a) An SEM image of the AlN piezo-optomechanical micro-disk resonator with a radius of 12\,$\mu$m. An angled zoomed-in view of the suspended disk is shown on the right. (b) An optical image of the frequency-tunable superconducting ``Ouroboros" microwave resonator fabricated in NbN on sapphire substrate. (c) An optical image of the integrated converter showing the alignment between the ``Ouroboros" and the micro-disk. The ``Ouroboros" is flipped over with its circular capacitor pad aligned above the micro-disk (indicated in a red dashed circle). Scale bars in (a)-(c) are 20, 100, 50\,$\mu\textrm{m}$, respectively. (d)-(f) Spectrum of the optical, the mechanical, and the microwave resonances, respectively, with both amplitude and phase response at 900\,mK. Orange lines are Lorentzian fittings. The detuning in (d) is from a pump laser at 194.363\,THz (1543.5\,nm). The inset in (f) shows the resonant frequency tuning of the ``Ouroboros" under an external magnetic field. The resonance in (f) is at zero field.}
\label{fig2}
\end{figure*}

The cavity piezoelectric interaction is characterized by a linear coupling rate $g_\mathrm{em}$, determined by the overlap between the microwave and the mechanical modes \cite{Han_2016}. For M-O photon conversion, an optical pump tone is needed to compensate the photon energy difference. As illustrated in Fig.\,\ref{fig1}(c), when the pump (purple arrow) is red-detuned from the optical resonance by $\sim$\,$\omega_\mathrm{m}$, the upper-sideband photon scattering (Stokes) is significantly enhanced by the optical mode so that efficient coupling between optical photons and phonons can be obtained with a linearized coupling rate given by $g_\mathrm{om}=\sqrt{n_\mathrm{cav}}g_\mathrm{om,0}$. Here, $g_\mathrm{om,0}$ is the single-photon coupling rate and $n_\mathrm{cav}$ is the intra-cavity photon number populated by the pump. Combining with the cavity electromechanical coupling, efficient bidirectional conversion between microwave ($\sim$\,10\,GHz) and optical photons ($\sim$\,200\,THz) can be achieved.

In the resolved-sideband limit ($\omega_\mathrm{m}\gg\kappa_\mathrm{o}$), the pump induced parametric amplification (anti-Stokes) can be neglected, and hence the system Hamiltonian can be expressed as (in the pump rotating frame)
\begin{equation}
\begin{split}
H/\hbar=&-\Delta_\mathrm{o}{\hat{a}^\dagger}\hat{a}+\omega_\mathrm{m}{\hat{b}^\dagger}\hat{b}+\omega_\mathrm{e}{\hat{c}^\dagger}\hat{c}\\
&-g_\mathrm{om}(\hat{a}^\dagger\hat{b}+\hat{b}^\dagger\hat{a})-g_\mathrm{em}(\hat{b}^\dagger\hat{c}+\hat{c}^\dagger{\hat{b}}),
\end{split}
\label{eq1}
\end{equation}
where $\Delta_\mathrm{o}\equiv\omega_\mathrm{p}-\omega_\mathrm{o}$ is detuning of the pump frequency $\omega_\mathrm{p}$ from the optical resonance. When the pump is detuned by $\Delta_\mathrm{o}=-\omega_\mathrm{m}$, the maximum photon number conversion efficiency can be obtained (at $\omega=\omega_\mathrm{m}$) \cite{SM}
\begin{equation}
\eta_0=\frac{\kappa_\mathrm{e,c}}{\kappa_\mathrm{e}}\frac{\kappa_\mathrm{o,c}}{\kappa_\mathrm{o}}\frac{4C_\mathrm{em}C_\mathrm{om}}{(C_\mathrm{em}+C_\mathrm{om}+1)^2+4\frac{(C_\mathrm{om}+1)^2}{\kappa_\mathrm{e}^2}\delta_\mathrm{em}^2}.
\label{eq2}
\end{equation}
Here, $C_\mathrm{em}\equiv\frac{4g_\mathrm{em}^2}{\kappa_\mathrm{e}\kappa_\mathrm{m}}$ and $C_\mathrm{om}\equiv\frac{4g_\mathrm{om}^2}{\kappa_\mathrm{o}\kappa_\mathrm{m}}$ are the electromechanical and the optomechanical cooperativities, respectively. $\delta_\mathrm{em}\equiv\omega_\mathrm{e}-\omega_\mathrm{m}$ is the frequency difference between the mechanical and the microwave resonances. To achieve the highest efficiency, ideally, the microwave and the mechanical modes should be on resonance ($\delta_\mathrm{em}=0$), and large and matched cooperativities ($C_\mathrm{em}=C_\mathrm{om}\gg1$) with very over-coupled readout ports ($\frac{\kappa_\mathrm{e,c}}{\kappa_\mathrm{e}},\frac{\kappa_\mathrm{o,c}}{\kappa_\mathrm{o}}\rightarrow1$) are desired. In a POM system, large optomechanical cooperativity has been previously demonstrated \cite{Chan2011,Forsch2019}. However, so far the achieved electromechanical cooperativity is only about $4\times10^{-3}$ \cite{Jiang2019} due to the lack of microwave cavity enhancement, which has become the main efficiency bottleneck of POM converters.

\section{Triple-resonance integration}

\begin{figure*}
\begin{centering}
\includegraphics{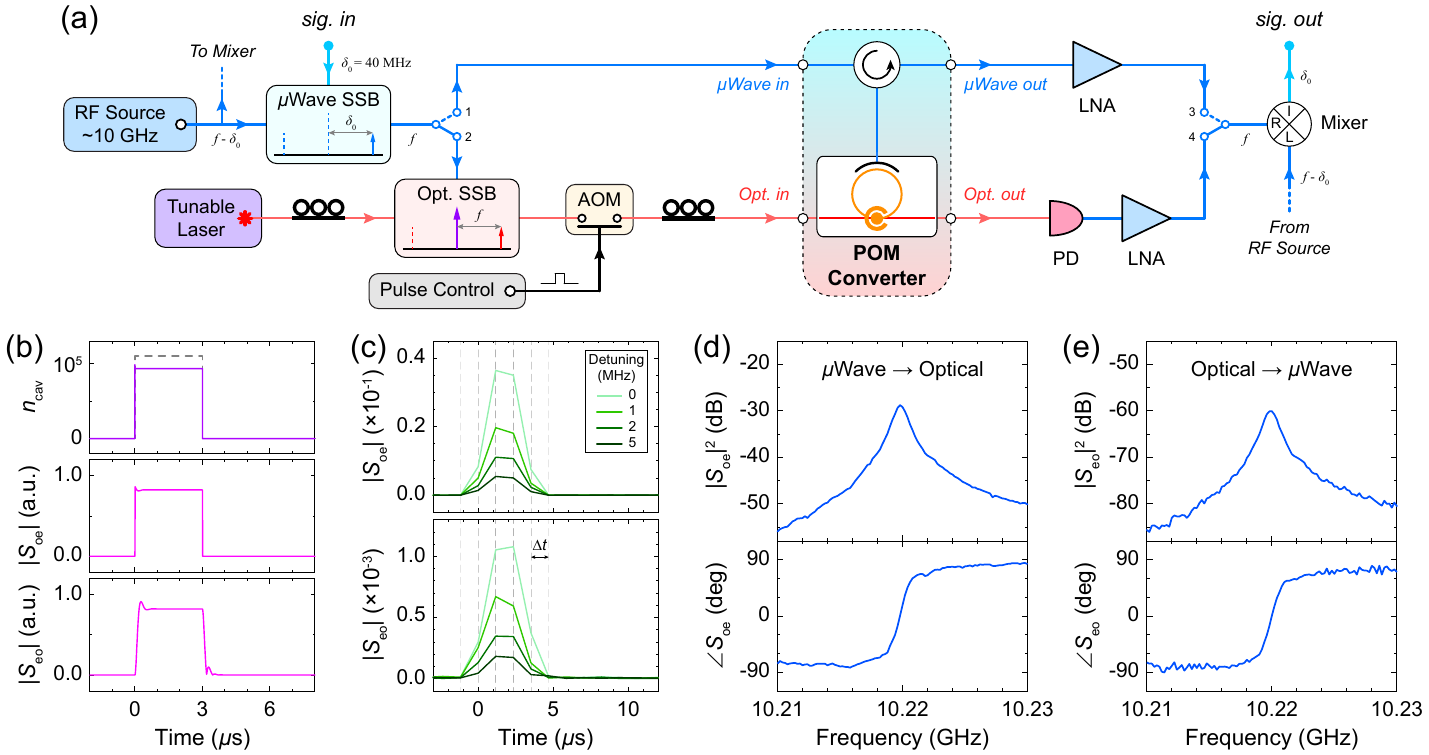}
\par\end{centering}
\caption{(a) A schematic of measurement scheme with pulsed optical pump. A 3-$\mu$s-wide red-detuned pump pulse is generated by on/off switching using an acoustic-optic modulator (AOM). To characterize the device temporal response, a low-frequency microwave signal ($\delta_0=40\,\textrm{MHz}$) from a lock-in amplifier (not shown here) is upconverted to $f$\,$\sim$\,10\,GHz around the mechanical resonance by single-sideband (SSB) modulation. The signal is then sent to the device either as the microwave input (switch pos.\,1), or the optical input after the optical SSB generation (pos.\,2). The microwave or optical output (pos.\,3 and 4, respectively) is then downconverted to $40\,\textrm{MHz}$ and sent to the lock-in amplifier to measure the quadratures in time domain. PD: photodetector. LNA: low-noise RF amplifier. (b) Theoretical calculation of the temporal response of the system under a pulsed pump (profile indicated in gray dashed line in the top panel). $n_\mathrm{cav}$: intra-cavity photon number. $S_\mathrm{oe}$ and $S_\mathrm{eo}$: microwave-to-optical and optical-to-microwave conversion signal, respectively. (c) Experimental data of the pulsed conversion signals at different input detunings from the mechanical resonance ($\frac{\omega_\mathrm{m}}{2\pi}=10.22\,\textrm{GHz}$). The measurement time resolution is ${\Delta}t\approx1.17\,\mu$s. (d) and (e) Coherent bidirectional microwave-optical photon conversion spectra with both amplitude and phase responses.}
\label{fig3}
\end{figure*}

We address this challenge and experimentally realize the triply resonant system by integrating a POM micro-disk and a planar superconducting microwave resonator as depicted in Fig.\,\ref{fig1}(d). The micro-disk simultaneously supports a high-quality ($Q$) factor optical whispering gallery mode and a high-frequency mechanical thickness mode at 10\,GHz \cite{Han2015}. To maximize the electromechanical mode overlap and coupling, we implement our frequency-tunable ``Ouroboros" design of the superconducting resonator \cite{Xu2019}. The ``Ouroboros" (yellow) forms a planar LC resonator which concentrates the electric field around its capacitor pads. Aligning the micro-disk in the close vicinity under the capacitor pad of the ``Ouroboros" allows maximized overlap between the perpendicular electric field of the microwave mode and the strain field of the mechanical thickness mode. As a result, cavity-enhanced electromechanical coupling is achieved. For converter operation, optical photons are sent into and read out from the micro-disk through an on-chip coupling waveguide; microwave input/output photons are inductively coupled to the ``Ouroboros" by using a loop probe.

An SEM image of the micro-disk is shown in Fig.\,\ref{fig2}(a), fabricated in aluminum nitride (AlN) with a thickness of 550\,nm and a radius of 12\,$\mu$m. The micro-disk is suspended and supported by a silicon dioxide anchor on top of a high-resistivity silicon substrate. The size of the anchor tip is minimized ($<$\,100\,nm in radius) by precise fabrication control for reducing acoustic radiation loss. Figure\,\ref{fig2}(b) shows an optical image of the superconducting ``Ouroboros" resonator fabricated in a 50-nm-thick niobium nitride (NbN) film on a thin sapphire substrate. In order to better align resonant frequency of the ``Ouroboros" with the thickness mode of the micro-disk, hole array structure is fabricated in the inductor wire of the ``Ouroboros" to realize frequency tunability by modifying its kinetic inductance via an external magnetic field \cite{Xu2019}. After separate chip fabrication, the ``Ouroboros" is flipped over and aligned above the micro-disk with approximately 5-$\mu$m spacing (Fig.\,\ref{fig2}(c)). The two chips are then bonded together, and two optical fibers are aligned and attached to the side-coupled optical waveguide of the micro-disk using epoxy. Device fabrication details are described in \cite{SM}.

The integrated superconducting POM device is loaded in a dilution refrigerator on the STILL flange (900\,mK) for measurement. Although the $\sim$\,1\,K environment presents slightly higher thermal noise ($\bar{n}_\mathrm{th}\approx1.6$), it provides much larger cooling power and thermal conductivity compared with millikelvin environment, which allows higher optical pump power for efficient photon conversion. Moreover, it is possible to radiatively cool a gigahertz mode and suppress its thermal occupancy for quantum operations \cite{Zhong2019} in spite of a hotter physical temperature of the resonator. We have recently demonstrated such radiative cooling of an ``Ouroboros" mode by constructing a cooling channel to a millikelvin cold bath \cite{Xu_cool,Wang2019}. In this work, we focus on the demonstration of coherent M-O photon conversion in our triply resonant POM system; systematic characterization of the noise performance as well as the implementation of radiative cooling is subject to future investigation.

\begin{figure*}
\begin{centering}
\includegraphics{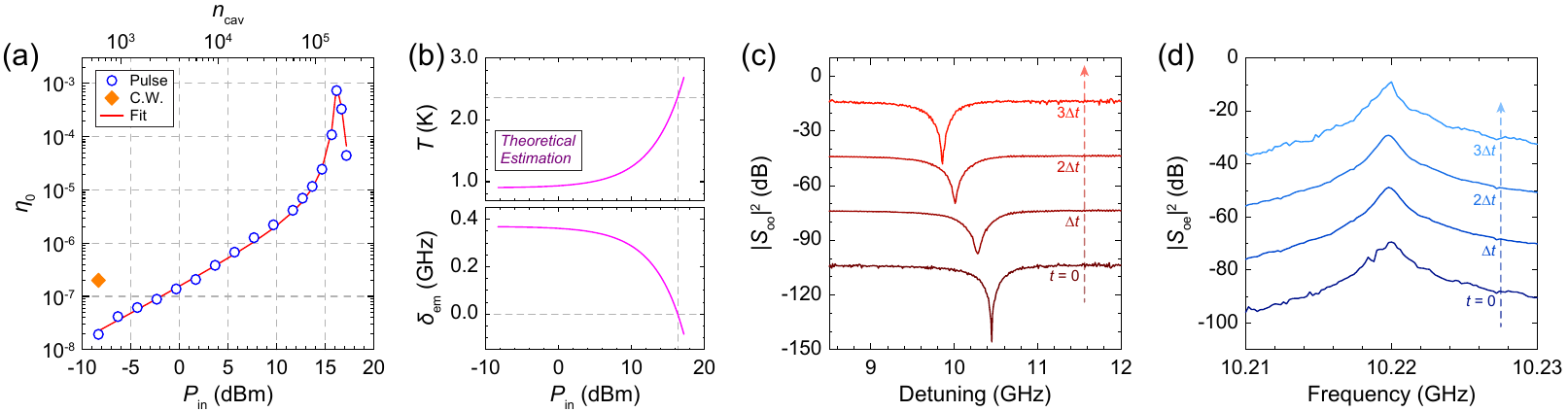}
\par\end{centering}
\caption{(a) Calibrated on-chip microwave-optical photon conversion efficiency $\eta_0$ at different optical pump power $P_\mathrm{in}$. (b) Estimated temperature $T$ and electromechanical resonance mismatch $\delta_\mathrm{em}$ of the ``Ouroboros" at different pump powers. (c) and (d) The optical resonance and the conversion spectrum at different time during a 16-dBm pump pulse, respectively. The spectra are vertically displaced for comparison purpose. The time resolution is ${\Delta}t\approx1.17\,\mu$s.}
\label{fig4}
\end{figure*}

The spectra of the optical, the mechanical, and the microwave resonances of the hybrid device are shown in Fig.\,\ref{fig2}(d), (e), and (f), respectively, measured at 900\,mK. The optical resonance is characterized using a single-sideband modulation scheme to reveal the phase response (details are explained later). The fiber-to-fiber optical transmission efficiency is measured to be 4.3\%, where the 14-dB loss mainly comes from the imperfect mode matching at the fiber-to-chip interfaces. The optical intrinsic and coupling $Q$ factors are fitted to be $Q_\mathrm{o,i}=5.4\times10^5$ and $Q_\mathrm{o,c}=5.8\times10^5$, respectively, corresponding to $\frac{\kappa_\mathrm{o,i}}{2\pi}=0.36\,\textrm{GHz}$ and $\frac{\kappa_\mathrm{o,c}}{2\pi}=0.34\,\textrm{GHz}$. The mechanical resonance of the micro-disk is characterized by electro-optomechanical driven response---similar to our previous room-temperature measurement scheme described in Ref.\,\cite{Han2015}. The thickness mode of the micro-disk is observed at $\frac{\omega_\mathrm{m}}{2\pi}=10.220\,\textrm{GHz}$ with $Q_\mathrm{m}=1.1\times10^4$ ($\frac{\kappa_\mathrm{m}}{2\pi}=0.93\,\textrm{MHz}$). The ``Ouroboros" resonance at zero external magnetic field is measured to be $\frac{\omega_\mathrm{e}}{2\pi}=10.778\,\textrm{GHz}$ with $Q_\mathrm{e,i}=2.6\times10^3$ ($\frac{\kappa_\mathrm{e,i}}{2\pi}=4.1\,\textrm{MHz}$) and $Q_\mathrm{e,c}=9.9\times10^4$ ($\frac{\kappa_\mathrm{e,c}}{2\pi}=0.11\,\textrm{MHz}$). This under-coupling of the microwave port is caused by the nonideal faraway position of the loop probe from the ``Ouroboros". By applying an external magnetic field, more than 200\,MHz frequency tuning range of the ``Ouroboros" resonance is realized without obvious degradation of the quality factor (inset in Fig.\,\ref{fig2}(f)). The single-photon optomechanical coupling rate is fitted to be $\frac{g_\mathrm{om,0}}{2\pi}\approx19\,\textrm{kHz}$ (details explained in Section \ref{EffSft}). Benefited from the triple-resonance design, a high electromechanical coupling rate $\frac{g_\mathrm{em}}{2\pi}\approx2.7\,\textrm{MHz}$ is achieved \cite{SM}, resulting in a large cooperativity $C_\mathrm{em}\sim7$.

\section{Pulsed photon conversion}

To realize efficient M-O photon conversion, a large optomechanical cooperativity is equally important. 
We exploit a pulsed-pump scheme to reduce the heating effect \cite{Forsch2019,Meenehan2015} and boost the intra-cavity photon number $n_\mathrm{cav}$. The measurement configuration is illustrated in Fig.\,\ref{fig3}(a). A continuous-wave (c.w.) pump tone (purple arrow) is provided by a stable tunable laser, red-detuned by $\Delta_\mathrm{o}=-\omega_\mathrm{m}$ from the optical resonance at 1543.420\,nm. The pump light is then pulsed by on/off switching using an acoustic-optic modulator (AOM). In order to characterize the coherent response of the POM converter in time domain, a low-frequency ($\delta_0=40\,\textrm{MHz}$) signal (labeled as \textit{sig.\,in} in Fig.\,\ref{fig3}(a)) from a lock-in amplifier is upconverted to $\sim$\,10\,GHz via microwave single-sideband (SSB) modulation technique. The upconverted c.w. microwave probe tone (blue arrow) can either be directly sent to the converter (switch pos.\,1) as its microwave input, or generate the optical input (red arrow) through an optical SSB modulator (switch pos.\,2). 
The microwave or the optical output (switch pos.\,3 or 4) is then downconverted to 40\,MHz and sent back to the lock-in amplifier for amplitude and phase measurement in time domain. Detailed experimental setup is explained in \cite{SM}.

Due to the finite temporal resolution of the lock-in amplifier (${\Delta}t\approx1.17\,\mu\textrm{s}$), we set a minimum pump pulse width of 3\,$\mu$s with a repetition period of 1\,ms. Theoretical analysis indicates that, within the pump pulse, our system already reaches a steady state. Numerical calculations of the system response are plotted in Fig.\,\ref{fig3}(b). The top panel shows the rapid population of intra-cavity optical photons under an ideal 3-$\mu$s pump; the fast response time corresponds to the optical cavity decay rate $1/\kappa_\mathrm{o}\approx0.2\,\textrm{ns}$. In experiment, the rise-/fall-time of the pump pulse and $n_\mathrm{cav}$ is determined by the AOM switching speed ($<$\,35\,ns), which is still much shorter than the pump pulse width.
The middle and the bottom panels show the microwave-to-optical ($S_\mathrm{oe}$) and the optical-to-microwave ($S_\mathrm{eo}$) conversion signal, respectively. In both cases, the transient response time is less than a few hundreds of nanoseconds, which is determined by the coupling and the dissipation rates $g_\mathrm{om},g_\mathrm{em},\kappa_\mathrm{e},\kappa_\mathrm{m}\sim\textrm{MHz}\times2\pi$ (see \cite{SM} for analysis). 

Figure\,\ref{fig3}(c) shows typical temporal M-O conversion pulses measured in experiment at different input detunings from the mechanical resonance. It can be seen that, when the pump is turned on, conversion signals are clearly observed in both directions with the maximum magnitude at zero detuning. By sweeping the input frequency, bidirectional M-O photon conversion spectra are obtained in Fig.\,\ref{fig3}(d) and (e) with both amplitude and phase responses, confirming the coherence of the conversion process. A large conversion bandwidth of 1\,MHz is achieved around the center mechanical resonant frequency.

\section{Conversion efficiency and thermal shift} 
\label{EffSft}

The photon conversion efficiency can be calibrated by measuring the full spectra of the scattering matrix elements. Here, we use  $S_{ij}[\omega]$ $(i,j=\mathrm{o,e})$ to denote the \textit{directly} measured reflection or conversion spectra, which are proportional to intrinsic scattering matrix elements $t_{ij}[\omega]$ of the converter (within the dashed box in Fig.\,\ref{fig3}(a)) up to a constant gain or loss factor. Since the system is reciprocal, the gain and loss factors along each input and output path can be calibrated out together to obtain the conversion efficiency $\eta[\omega]=\abs{t_\mathrm{oe}[\omega]}^2=\abs{t_\mathrm{eo}[\omega]}^2$ \cite{SM}, which simplifies to Eq.\,(\ref{eq2}) at the peak of the spectrum ($\eta[\omega_\mathrm{m}]=\eta_0$). The on-chip M-O photon conversion efficiencies $\eta_0$ at different optical pump powers $P_\mathrm{in}$ are plotted as blue circles in Fig.\,\ref{fig4}(a). The corresponding intra-cavity photon number $n_\mathrm{cav}$, which is proportional to $P_\mathrm{in}$, is labeled on the upper $x$-axis. It can be seen that at low powers, $\eta_0$ increases linearly with $P_\mathrm{in}$ and $n_\mathrm{cav}$, as expected from Eq.\,(\ref{eq2}) when $C_\mathrm{om}\ll1$. As the pump power further increases, the efficiency dramatically goes up to a peak value and then drops. This behavior can be attributed to the pump induced thermal shift of the microwave resonance as discussed below. The highest efficiency achieved is $(7.3\pm0.2)\times10^{-4}$, which corresponds to an internal efficiency of $(5.8\pm0.2)\%$ after excluding the low extraction factor ($\frac{\kappa_\mathrm{o,c}}{\kappa_\mathrm{o}}\frac{\kappa_\mathrm{e,c}}{\kappa_\mathrm{e}}\approx1.3\%$).
 
The thermal effect can be observed from the optical resonance shift during the pump pulse. Figure\,\ref{fig4}(c) shows the optical spectrum at different time within the pulse at a pump power of 16\,dBm. It can be seen that after the pump is turned on, the optical resonance gradually shifts to lower frequencies by $\sim$\,0.59\,GHz within the 3-$\mu$s pulse. Therefore, in our experiment, the detuning of the optical pump is always adjusted and optimized especially at high powers to compensate the thermal shift and achieve the highest conversion efficiency.
%Using the thermal expansion coefficient of AlN $c=\question{??}$ at low temperature, we estimated the temperature increase of the micro-disk to be $??$. 
In contrast, the thermal effect has negligible influence on the mechanical resonance since the small relative frequency shift ($<10^{-5}$) corresponds to less than $0.1\,\textrm{MHz}$ at the mechanical frequency, which is much smaller than the mechanical linewidth and hence not noticeable in the conversion spectrum (Fig.\,\ref{fig4}(d)).

In our POM system, the ``Ouroboros" has an original resonance at 10.778\,GHz, higher than the mechanical resonance of the micro-disk at 10.220\,GHz. Therefore, an external magnetic field ($\sim$\,1\,mT) is applied to tuned the ``Ouroboros" frequency down to 10.590\,GHz to reduce the frequency mismatch. At low optical pump powers, the remained mismatch ($\delta_\mathrm{em}$ in Eq.\,\ref{eq2}) compromises the electromechanical interaction and hence limits the conversion efficiency. Nevertheless, at high pump powers, the heating effect (mostly comes from the fiber-to-chip interface in our case) becomes non-negligible, and the ``Ouroboros" frequency will further decrease due to the reduction of Cooper pair density in the superconductor. As a result, raising the pump power will not only boost the optomechanical cooperativity $C_\mathrm{om}$, but also reduce the electromechanical resonance mismatch $\delta_\mathrm{em}$, thus giving rise to a sharp increase of the conversion efficiency.

Due to the under-coupling condition of the microwave probe and the imperfect measurement background, it is difficult to trace the ``Ouroboros" resonance at high pump powers. However, we can estimate the frequency shift and the temperature of the ``Ouroboros" based on the pump power dependence of the conversion efficiency. Because the inductance of the ``Ouroboros" is dominated by the kinetic inductance ($L_\mathrm{k}$), we assume that its resonance frequency shift is affected by temperature primarily via $\omega_\mathrm{e}(T)\propto\frac{1}{\sqrt{L_\mathrm{k}(T)}}$. Using the temperature dependence of kinetic inductance from BCS theory \cite{Annunziata2010}, we have
\begin{equation}
\omega_\mathrm{e}(T)\approx\nu_0\Bigg[\sqrt{1-\frac{T}{T_\mathrm{c}}}\tanh{\left(1.53\frac{T_\mathrm{c}}{T}\sqrt{1-\frac{T}{T_\mathrm{c}}}\right)}\Bigg]^{1/2},
\label{eq3}
\end{equation}
where $T_\mathrm{c}\approx12$\,K is the critical temperature of the NbN film, and $T\approx{T_0}+\beta P_\mathrm{in}$ assuming that the temperature change from $T_0=900$\,mK is proportional to the pump power $P_\mathrm{in}$. $\nu_0$ and $\beta$ are two proportional constants. $\nu_0$ can be determined by the initial Ouroboros frequency $\frac{\omega_\mathrm{e}(T)}{2\pi}=10.590$\,GHz at $T=T_0$, while $\beta$ is treated as a fitting parameter.

Substituting Eq.\,(\ref{eq3}) into Eq.\,(\ref{eq2}), we can get the power dependence of conversion efficiency. By setting $\beta$, $g_\mathrm{om,0}$, and $g_\mathrm{em}$ as free parameters while using experimentally obtained values for other device parameters, the conversion efficiency can be fitted very well as indicated in the red line in Fig.\,\ref{fig4}(a). The thermal constant and the single-photon optomechanical coupling rate are extracted to be $\beta=(3.43\pm0.02)\times10^{-2}$\,K/mW and $\frac{g_\mathrm{om,0}}{2\pi}=(19\pm2)$\,kHz, respectively. The electromechanical coupling rate is fitted to be $\frac{g_\mathrm{em}}{2\pi}=(3.2\pm0.3)$\,MHz, which agrees well with our experimental characterization ($\sim$\,2.7\,MHz, see \cite{SM}).
The corresponding temperature and resonant frequency shift of the ``Ouroboros" at different pump powers are plotted in Fig.\,\ref{fig4}(b). As expected, the temperature gradually goes up, whereas the resonant frequency decreases, with increasing pump power. At the highest conversion efficiency, the fitting indeed infers zero electromechanical resonance mismatch (gray dashed lines in Fig.\,\ref{fig4}(b)). The theoretical estimation also indicates that the ``Ouroboros" temperature remains below $\sim$\,3\,K even at the highest pump power that we applied. This is reasonable since the thermally induced relative frequency shift of the superconducting ``Ouroboros" is only $\sim$\,4\%.
It is worth pointing out that the measured efficiency-pump power dependence in Fig.\,\ref{fig4}(a) is highly repeatable without observation of hysteresis effect (see \cite{SM} for details). This confirms that during our measurement the ``Ouroboros" always remains in its superconducting state with its temperature well below $T_\mathrm{c}$, even in presence of the external magnetic bias and the pump induced heating.

Therefore, by implementing the pulsed pump scheme, we are able to significantly reduce the optical heating and boost the pump photon number. The optomechanical cooperativity at the highest conversion efficiency is estimated to be $C_\mathrm{om}\sim0.4$. For comparison, we also perform measurement using c.w. optical pump. Affected by excessive heating, the best c.w. conversion efficiency obtained is only $2\times10^{-7}$ (orange diamond point in Fig.\,\ref{fig4}(a)) with the pump power limited to $-8.3$\,dBm, which is over two orders of magnitude less than that can be applied in the pulsed scheme.

\section{Discussion} 
Benefited from the integration of the superconducting microwave cavity, we demonstrated a large $C_\text{em}\sim7$ in our triply resonant POM interface. To further improve the conversion efficiency, $C_\mathrm{om}$ needs to be enhanced to match $C_\mathrm{em}$. This could be done by either improving the optical and the mechanical quality factors through better fabrication process or further increasing the pump photon number. The pump induced heating would be reduced if the fiber-to-chip coupling loss is minimized.

Towards quantum-enabled M-O photon conversion, it is equally important to minimize the added noise induced by thermal excitations. The high operation frequency of our converter is advantageous to suppress thermal noise from both microwave and mechanical baths. The estimated temperature of the ``Ouroboros" leads to a thermal occupancy of the microwave bath around only a few photons ($\bar{n}_\mathrm{the}\approx5.6$ at 3\,K). In \cite{SM}, we provide comprehensive theoretical analysis of the added noise during the photon conversion process in both directions. 
We show that it is possible to suppress both the microwave and the mechanical added noise to subphoton level even the device is physically in a relatively ``hot" environment. This can be realized by implementing the radiative cooling scheme \cite{Xu_cool,Wang2019} with an over-coupled microwave port.
In addition, a large $C_\mathrm{om}$ is needed to suppress the mechanical added noise during the optical-to-microwave photon conversion. Future experimental characterization of the noise performance of our POM converter will be valuable.

In summary, we have realized an integrated interface between superconducting and nanophotonic circuits. The demonstrated high-frequency superconducting cavity POM system provides a new route towards the development of scalable efficient M-O photon conversion devices. Combining with our recently demonstrated radiative cooling technique and feasible future improvement of device parameters, quantum operations braiding superconducting microwave, nanomechanics, and nanophotonics can be expected. 

$\\$

\begin{acknowledgments}
This work is supported by ARO grant W911NF-18-1-0020. The authors also acknowledge partial supports from AFOSR MURI (FA9550-14-1-0052, FA9550-15-1-0015), DOE (DE-SC0019406), NSF (EFMA-1640959) and the Packard Foundation. The authors thank M. Power and M. Rooks for assistance in device fabrication. 
\end{acknowledgments}

%merlin.mbs apsrev4-1.bst 2010-07-25 4.21a (PWD, AO, DPC) hacked
%Control: key (0)
%Control: author (72) initials jnrlst
%Control: editor formatted (1) identically to author
%Control: production of article title (0) allowed
%Control: page (0) single
%Control: year (1) truncated
%Control: production of eprint (-1) disabled
%

%%%%%%%%%%%%%%%%%%%%%%%%%%%%%%%%%%%%%%%%%%%%%%%%%%%%%%%%%%%%%%%%%%%
%%%%%%%%%%%%%%%%%%%      Supplementary      %%%%%%%%%%%%%%%%%%%%%%%
%%%%%%%%%%%%%%%%%%%%%%%%%%%%%%%%%%%%%%%%%%%%%%%%%%%%%%%%%%%%%%%%%%%

\onecolumngrid
\newpage
\clearpage

\begin{spacing}{1.3}
\begin{center}
  \textbf{\large Supplemental Material for ``{Cavity piezo-mechanics for superconducting-nanophotonic quantum interface}''}
\end{center}
\end{spacing}

\beginsup

\section{Device fabrication}
\label{appfab}
The POM micro-disk is fabricated in a 550-nm-thick \textit{c}-axis aligned AlN layer on an oxidized high-resistivity silicon substrate. A 200-nm-thick silicon dioxide (SiO$_2$) layer is first deposited on top of the AlN as hard mask via plasma-enhanced chemical vapor deposition (PECVD). The photonic pattern is then defined using electron-beam lithography (EBL) with Ma-N\,2403 resist, followed by resist reflow technique to improve surface smoothness. Fluorine-based reactive-ion etching (RIE) is performed to transfer the pattern in the SiO$_2$ mask layer. The AlN is then shallow-etched using chlorine-based RIE with a thin layer ($\sim$\,100\,nm) left as the protection layer during the later releasing process. The second EBL is performed with ZEP\,520A resist to define the releasing areas where the rest of the AlN is etched through by the second chlorine-based RIE. In order to integrate with the ``Ouroboros" chip, PECVD SiO$_2$ and silicon nitride (SiN) layers (5\,$\mu$m and 500\,nm thick, respectively) are deposited successively around edge areas of the AlN chip as spacers besides the micro-disk. The SiN protects the SiO$_2$ spacers from being etched during the releasing process. The device is then released in buffered oxide etch (BOE) to etch the sacrificial SiO$_2$ layer beneath the AlN and suspend the micro-disk. The releasing process is precisely timed so that the SiO$_2$ anchor of the micro-disk has a minimized tip size around 100\,nm in radius. The chip is finally cleaved to prepare for fiber side coupling at the two ends of the photonic waveguide.

The superconducting ``Ouroboros" is fabricated in a 50-nm-thick NbN film deposited on a 150-$\mu$m-thick sapphire substrate. The ``Ouroboros" structure is defined by EBL using hydrogen silsesquioxane (HSQ) resist. After individual fabrication and characterization of the micro-disk and the ``Ouroboros", the two chips are aligned under a microscope and integrated together. The ``Ouroboros" chip is flipped over and placed on top of the AlN chip with the circular capacitor pad of the ``Ouroboros" aligned above the AlN micro-disk. The SiO$_2$ spacers maintain a $\sim$\,5-$\mu$m gap between the ``Ouroboros" and the micro-disk so that optics and superconductivity would not compromise each other. This gap, on the other hand, is still much smaller than the microwave wavelength and mode size, which ensures large piezo-electromechanical coupling. The aligned chips are then bonded together by applying ultraviolet (UV) epoxy at the edges. Two ultra-high numerical aperture (UHNA) optical fibers are aligned to the two ends of the coupling waveguide of the micro-disk, then bonded using UV epoxy. The integrated superconducting POM device is then enclosed in a copper box holder for measurement, with a home-made superconducting coil placed above the ``Ouroboros" to provide magnetic field for frequency tuning.

%%%%%%%%%%%%%%%%%%%%%%%%%%%%%%%%%%%%%%%%%%%%%%%%%%%%%%%%%%%%%%%%%%%

\section{Electromechanical coupling rate, $\boldsymbol{g_\mathrm{em}}$}
\label{appgem}

The electromechanical coupling rate $g_\mathrm{em}$ can be characterized by measuring the microwave reflection spectrum of the coupled modes. Since in our converter device, the mechanical resonance of the micro-disk is out of the frequency tuning range of the ``Ouroboros", we experimentally characterize $g_\mathrm{em}$ using a separate device, fabricated in the same way as the converter but the ``Ouroboros" has slightly lower resonant frequency which can be aligned with mechanical thickness mode of the micro-disk. 

\begin{figure}
\begin{centering}
\includegraphics{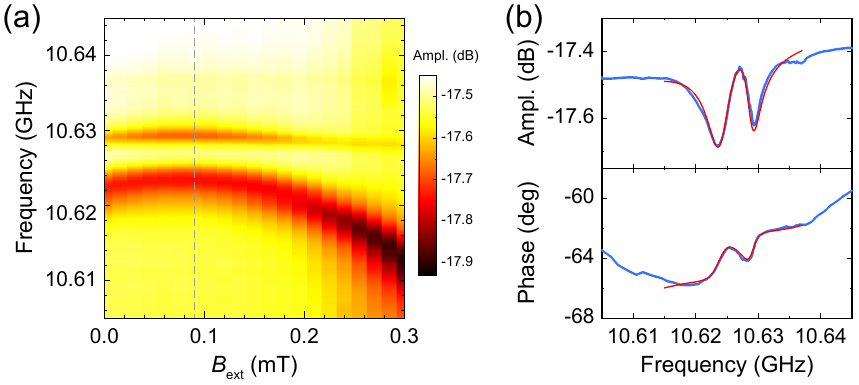}
\par\end{centering}
\caption{Measurement of the electromechanical coupling rate $g_{\mathrm{em}}$. (a) Microwave reflection spectrum at different external magnetic fields. As the ``Ouroboros" resonance is tuned across the mechanical resonance of the micro-disk, avoided crossing is observed. (b) Line plot of the spectrum at $B_\mathrm{ext}=0.09$\,mT (gray dashed line in (a)) with both amplitude and phase responses. The red lines are the fitting using coupled mode formula, which extracts a coupling rate $\frac{g_\mathrm{em}}{2\pi}=2.7$\,MHz.}
\label{figgem}
\end{figure}

As shown in Fig.\,\ref{figgem}(a), when the ``Ouroboros" resonance is tuned to lower frequencies by the external magnetic field, avoided crossing is observed, indicating strong electromechanical coupling. Figure\,\ref{figgem}(b) plots the amplitude and phase spectrum at $B_\mathrm{ext}=0.09$\,mT (gray dashed line in (a)), which can be fitted by using the microwave reflection coefficient without the optomechanical coupling
\begin{equation}
S_\mathrm{ee}[\omega]\Big\vert_{g_\mathrm{om}=0} =-1+\frac{\kappa_\mathrm{e,c}}{-i(\omega-\omega_\mathrm{e})+\frac{\kappa_\mathrm{e}}{2}+\frac{g_\mathrm{em}^2}{-i(\omega-\omega_\mathrm{m})+\frac{\kappa_\mathrm{m}}{2}}}.
\end{equation}
From the fitting (red lines in Fig.\,\ref{figgem}(b)), we extract $\frac{g_\mathrm{em}}{2\pi}=2.7$\,MHz. The intrinsic microwave and mechanical \textit{Q} factors and dissipation rates of this devices are $Q_\mathrm{e,i}=1.7\times10^3$ ($\frac{\kappa_\mathrm{e,i}}{2\pi}=6.4$\,MHz) and $Q_\mathrm{m}=1.4\times10^4$ ($\frac{\kappa_\mathrm{m}}{2\pi}=0.74$\,MHz).

It is worth pointing out that the slightly different resonant frequencies and \textit{Q} factors of the device have negligible influence on the microwave and the mechanical mode profiles and their overlap; hence won't affect the electromechanical coupling. As discussed in the main text, this directly measured $g_\mathrm{em}$ from the characterization device agrees very well with the fitted value based on the power dependence of conversion efficiency of our converter device. For a conservative estimation, we use $\frac{g_\mathrm{em}}{2\pi}=2.7$\,MHz and calculate $C_\mathrm{em}\approx7.4$.

%%%%%%%%%%%%%%%%%%%%%%%%%%%%%%%%%%%%%%%%%%%%%%%%%%%%%%%%%%%%%%%%%%%

\section{Measurement setup}
\label{appsetup}

\begin{figure}
\begin{centering}
\includegraphics{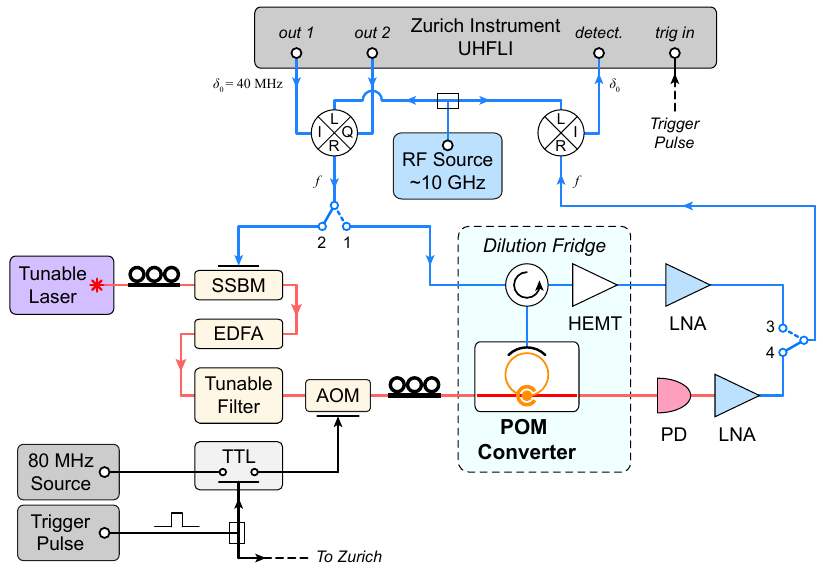}
\par\end{centering}
\caption{Measurement setup of the pulsed photon conversion. UHFLI: ultra-high-frequency lock-in amplifier. SSBM: single-sideband modulator. EDFA: erbium-doped fiber amplifier. AOM: acoustic-optic modulator. TTL: transistor-transistor logic switch. HEMT: high-electron-mobility transistor amplifier. PD: photodetector. LNA: low-noise RF amplifier.}
\label{figsetup}
\end{figure}

Detailed measurement setup of the pulsed microwave-optical photon conversion is shown in Fig.\,\ref{figsetup}. An ultra-high-frequency lock-in amplifier (Zurich Instruments UHFLI) is used to send two low-frequency ($\delta_0=40\,\textrm{MHz}$) signals to an IQ mixer, with their phase difference fine-tuned around $90^\circ$ for synthesizing microwave single sideband from an RF source at $\sim$\,10\,GHz as local oscillator (LO). This upconverted single-sideband microwave tone can be swept by varying the frequency of the LO, and sent to either the converter (switch pos.\,1) as a c.w. microwave input or an optical single-sideband modulator (SSBM) to generate an optical input (switch pos.\,2). The optical single sideband is generated from a stable tunable laser (Santec\,TSL-710), which at the same time serves as the pump after amplified by an erbium-doped fiber amplifier (EDFA) and filtered by a tunable filter to reduce noise. This c.w. pump is pulsed by switching on/off an acoustic-optic modulator (AOM). This is done by sending a 30-$\mu$s trigger pulse with a repetition period of 1\,ms from a function generator to a TTL switch to modulate the 80-MHz RF carrier of the AOM. At the same time, the trigger pulse is split and sent to the Zurich UHFLI as the external trigger for temporal measurement. The output signal of the converter at $\sim$\,10\,GHz is downconverted to 40\,MHz using the same RF source and sent to the Zurich for detection in time domain.

%%%%%%%%%%%%%%%%%%%%%%%%%%%%%%%%%%%%%%%%%%%%%%%%%%%%%%%%%%%%%%%%%%%

\section{Measurement repeatability and efficiency calibration}
\label{appcali}

\begin{figure}
\begin{centering}
\includegraphics{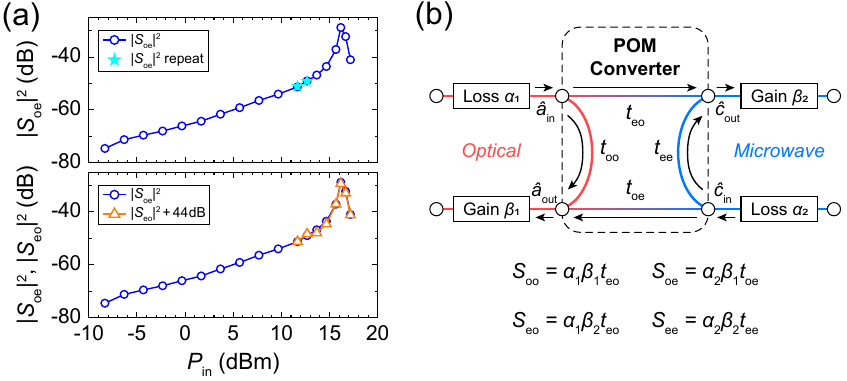}
\par\end{centering}
\caption{(a) Directly measured bidirectional conversion signals at different pump power $P_\mathrm{in}$. The upper panel shows repeated measurement of $S_\mathrm{oe}$ at lower powers after sweeping to the highest applied pump power. The results overlap very well with the original data without hysteresis effect. The lower panel shows the consistent power dependence of $S_\mathrm{oe}$ and $S_\mathrm{eo}$. The $\abs{S_\mathrm{eo}}^2$ curve is manually shifted up by 44\,dB for comparison purpose. (b) Signal flowing diagram showing the relation between the directly measured scattering matrix elements $S_{ij}$ and the intrinsic on-chip elements $t_{ij}$, $(i,j=\textrm{o,e})$. The gain and loss factors ($\alpha$'s and $\beta$'s) can be calibrated out together to obtain the intrinsic conversion efficiency.}
\label{figcali}
\end{figure}

The bidirectional conversion signals in our experiment are highly repeatable within the largest applied pump power at 17.2\,dBm. As indicated as cyan stars in upper panel of Fig.\,\ref{figcali}(a), we repeat the $S_\mathrm{oe}$ measurement at low powers after sweep up to highest pump power. It can be seen that even after a few hours of delay between measurements, the results still overlap very well with previous data, confirming no hysteresis effects. This indicates that during our experiment, the ``Ouroboros" remains well in its superconducting state with stable and repeatable resonant frequency even in presence of the pump induced heating and the external magnetic bias field. In addition, we plot the directly measured $S_\mathrm{eo}$ in the lower panel of Fig.\,\ref{figcali}(a) to show consistent power dependence as the $S_\mathrm{oe}$ ($\abs{S_\mathrm{eo}}^2$ is manually shifted by 44\,dB for comparison purpose), revealing the bidirectional and reciprocal nature of the M-O photon conversion.

The conversion efficiency is calibrated by measuring the full spectra of the scattering matrix elements, following the procedure described in \cite{Fan2018}. The principle of the calibration is illustrated in Fig.\,\ref{figcali}(b). The directly measured scattering matrix elements $S_{ij}$ are proportional to the intrinsic elements of the converter up to a constant gain or loss factor. Although it is difficult to calibrate individual gain or loss factor along each path, they can be canceled out together to obtain the intrinsic conversion efficiency since conversion is reciprocal. The on-chip efficiency (peak of the spectrum at $\omega=\omega_\mathrm{m}$) can be obtained by
\begin{equation}
\eta_0=\abs{t_\mathrm{oe}[\omega_\mathrm{m}]t_\mathrm{eo}[\omega_\mathrm{m}]}=\frac{\abs{S_\mathrm{oe}[\omega_\mathrm{m}]S_\mathrm{eo}[\omega_\mathrm{m}]}}{\abs{S_\mathrm{oo,bg}[\omega_\mathrm{m}]S_\mathrm{ee,bg}[\omega_\mathrm{m}]}},
\label{eqS3}
\end{equation}
where $S_\mathrm{oo,bg}$ and $S_\mathrm{ee,bg}$ denote the background of the reflection spectra $S_\mathrm{oo}$ and $S_\mathrm{ee}$ without resonance, namely, $\alpha_1\beta_1$ and $\alpha_2\beta_2$. By fitting the reflection and conversion spectra at the optimal pump power of 16.1\,dBm, we extract $\abs{S_\mathrm{oo,bg}[\omega_\mathrm{m}]}^2=(4.27\pm0.02)\times10^{-2}$, $\abs{S_\mathrm{ee,bg}[\omega_\mathrm{m}]}^2=(6.44\pm0.24)\times10^{-2}$, $\abs{S_\mathrm{oe,bg}[\omega_\mathrm{m}]}^2=(1.29\pm0.04)\times10^{-2}$, $\abs{S_\mathrm{eo,bg}[\omega_\mathrm{m}]}^2=(1.15\pm0.02)\times10^{-2}$. Using Eq.\,(\ref{eqS3}) and the propagation of uncertainty, we can calibrate the highest on-chip conversion efficiency to be $(7.3\pm0.2)\times10^{-4}$, corresponding to an internal efficiency of $(5.8\pm0.2)\%$.

%%%%%%%%%%%%%%%%%%%%%%%%%%%%%%%%%%%%%%%%%%%%%%%%%%%%%%%%%%%%%%%%%%%

\section{Pulse response and conversion spectrum}
\label{apppulse}

\subsection{Optical cavity photon response}
We first calculate the response of the classical intra-cavity field $\alpha(t)$ under a pulse pump input $\alpha_\mathrm{in}(t)$, which is related to the input pump power via $P_\mathrm{in}(t)=\hbar\omega_\mathrm{o}\abs{\alpha_\mathrm{in}(t)}^2$. To first order approximation, the optomechanical backaction can be neglected and $\alpha$ simply follows 
\begin{equation}
\Dot{\alpha}(t)=\Big(i\Delta_\mathrm{o}-\frac{\kappa_\mathrm{o}}{2}\Big)\alpha(t)+\sqrt{\kappa_\mathrm{o,c}}\alpha_\mathrm{in}(t).
\end{equation}
If the rising edge of the input is a perfect step function, namely, $\alpha_\mathrm{in}(t)=0$ when $t<0$ and $\alpha_\mathrm{in}(t)=\alpha_0$ is constant when $t\ge0$, the intra-cavity photon number can be solved as ($t\ge0$)
\begin{equation}
n_\mathrm{cav}(t)\equiv\abs{\alpha(t)}^2=\frac{\kappa_\mathrm{o,c}\abs{\alpha_0^2}}{\Delta_\mathrm{o}^2+\kappa_\mathrm{o}^2/4}\left[1+e^{-\kappa_\mathrm{o}t}-2\cos{(\Delta_\mathrm{o}t)}e^{-\frac{\kappa_\mathrm{o}}{2}t}\right],
\label{ncav}
\end{equation}
which reaches steady state with a time constant $1/\kappa_\mathrm{o}$. Similarly, the falling edge will have the same response time. The cosine oscillation term in solution Eq.\,(\ref{ncav}) is due to the perfectly sharp step edge of the input which doesn't exist in experiment. Therefore, in numerical calculation in Fig.\,\ref{fig3}(b) of the main text, we used a Gauss error function for a smoother input, instead of a step function, to reduce the artificial oscillation.

\subsection{Conversion signal response}
Now we analyze the pulse response of the conversion photon. Due to the fast response of the $n_\mathrm{cav}(t)$, the optomechanical coupling $g_\mathrm{om}(t)\equiv g_\mathrm{om,0}\sqrt{n_\mathrm{cav}(t)}$ can be simply treated as a step function. In other words, the pump pulse serves as a fast switch to quickly turn on and off the optomechanical coupling. In the resolved-sideband limit ($\omega_\mathrm{m}\gg\kappa_\mathrm{o}$), the Heisenberg equations of motion of the intra-cavity field $\mathbf{a}(t)=(\hat{a},\hat{c},\hat{b})^T$ can be written as \cite{Zhong2019}
\begin{equation}
\mathbf{\Dot{a}}(t)=\mathbf{A a}(t)+\mathbf{B a_{in}}(t),
\label{EOM}
\end{equation}
where $\mathbf{a_{in}}(t)=(a_\mathrm{in1},a_\mathrm{in2},c_\mathrm{in1},c_\mathrm{in2},b_\mathrm{in2})^T$ is the input term. The subscript (1, 2) denotes the coupling or the intrinsic (noise) port, respectively. Matrices $\mathbf{A}$ and $\mathbf{B}$ are given by
\begin{equation}
\begin{split}
\mathbf{A} =
\begin{pmatrix}
i\Delta_{\mathrm{o}}-\frac{\kappa_{\mathrm{o}}}{2}	&	0   &	ig_\mathrm{om}(t)\\
0	&	-i\omega_{\mathrm{e}}-\frac{\kappa_{\mathrm{e}}}{2}	&	ig_\mathrm{em}\\
ig_\mathrm{om}(t)	&	ig_\mathrm{em}	&	-i\omega_{\mathrm{m}}-\frac{\kappa_{\mathrm{m}}}{2}\\
\end{pmatrix},
\quad
\mathbf{B} =
\begin{pmatrix}
\sqrt{\kappa_{\mathrm{o,c}}}	&	\sqrt{\kappa_{\mathrm{o,i}}}	&	0 	&	0	&	0 \\
0	&	0	&	\sqrt{\kappa_{\mathrm{e,c}}}    &	\sqrt{\kappa_{\mathrm{e,i}}}	&	0 \\
0	&	0	&	0	&	0	&	\sqrt{\kappa_{\mathrm{m}}} \\
\end{pmatrix}.
\end{split}
\end{equation}
Combining with the input-output relation
\begin{equation}
\mathbf{a_{out}}(t)=\mathbf{B}^T\mathbf{a}(t)-\mathbf{a_{in}}(t)
\label{IO}
\end{equation}
and setting the noise input terms to be zero, we can numerically calculate the temporal profiles of the conversion signals as plotted in Fig.\,\ref{fig3}(b) of the main text.

The physical understanding of the transient conversion response is as follows. In general, characteristic response time of a cavity is determined by its total energy exchange rate (including all dissipation and coupling channels). For the mechanical mode, besides its own intrinsic dissipation $\kappa_\mathrm{m}$, the electromechanical coupling provides an effective energy transfer rate $\sim$\,$\frac{4g_\mathrm{em}^2}{\kappa_\mathrm{o}}=C_\mathrm{em}\kappa_\mathrm{m}$. When $g_\mathrm{om}$ is turned on by the pump pulse, the effective optomechanical energy transfer rate is $\sim$\,$C_\mathrm{om}\kappa_\mathrm{m}$. Therefore, the mechanical total energy exchange rate is $\Gamma_\mathrm{m}\sim(1+C_\mathrm{em}+C_\mathrm{om})\kappa_\mathrm{m}$, which results in a response time $\sim$\,$1/\Gamma_\mathrm{m}\ll1$\,$\mu$s. Similarly, the microwave total rate is $\Gamma_\mathrm{e}\sim(1+C_\mathrm{em})\kappa_\mathrm{e}$ and $1/\Gamma_\mathrm{e}\ll1$\,$\mu$s. Therefore, as confirmed by the numerical results, the conversion signals can quickly reach steady state during the 3-$\mu$s pump pulse.

\subsection{Conversion spectrum}
Solving Eq.\,(\ref{EOM}) and (\ref{IO}) in the frequency domain, we can define the scattering matrix $\mathbf{S}[\omega]$ from $\mathbf{a_{out}}[\omega]=\mathbf{S}[\omega]\mathbf{a_{in}}[\omega]$, which can be expressed as
\begin{equation}
\mathbf{S}[\omega]=\mathbf{B}^T[-i\omega\mathbf{I}_3-\mathbf{A}]^{-1}\mathbf{B}-\mathbf{I}_{5},
\end{equation}
where $\mathbf{I}_{3}$ and $\mathbf{I}_{5}$ are $3\times3$ and $5\times5$ identity matrix, respectively. The (1,3) and the (3,1) elements of $\mathbf{S}$ correspond to the microwave-to-optical and the optical-to-microwave conversion coefficients ($t_\mathrm{oe}$ and $t_\mathrm{eo}$ in the main text), respectively; namely, $\mathrm{S_{1,3}}[\omega]\equiv a_\mathrm{out1}[\omega]/c_\mathrm{in1}[\omega]=t_\mathrm{oe}[\omega]$ and $\mathrm{S_{3,1}}[\omega]\equiv c_\mathrm{out1}[\omega]/a_\mathrm{in1}[\omega]=t_\mathrm{eo}[\omega]$. The M-O photon conversion spectrum $\eta[\omega]\equiv\abs{t_\mathrm{oe}[\omega]}^2=\abs{t_\mathrm{eo}[\omega]}^2$ can be obtained as
%\begin{widetext}
\begin{equation}
\eta[\omega]=\frac{\kappa_\mathrm{e,c}}{\kappa_\mathrm{e}}\frac{\kappa_\mathrm{o,c}}{\kappa_\mathrm{o}}\frac{4C_\mathrm{em}C_\mathrm{om}}{\abs{C_\mathrm{em}\big(1-i\frac{\omega+\Delta_\mathrm{o}}{\kappa_\mathrm{o}/2}\big)+\big(1-i\frac{\omega-\omega_\mathrm{e}}{\kappa_\mathrm{e}/2}\big)\Big[C_\mathrm{om}+\big(1-i\frac{\omega+\Delta_\mathrm{o}}{\kappa_\mathrm{o}/2}\big)\big(1-i\frac{\omega-\omega_\mathrm{m}}{\kappa_\mathrm{m}/2}\big)\Big]}^2},
\end{equation}
%\end{widetext}
which reduces to Eq.\,(\ref{eq2}) in the main text when $\Delta_\mathrm{o}=-\omega_\mathrm{m}$ and $\omega=\omega_\mathrm{m}$.

%%%%%%%%%%%%%%%%%%%%%%%%%%%%%%%%%%%%%%%%%%%%%%%%%%%%%%%%%%%%%%%%%%%

\section{Noise analysis}
\label{appnoise}

The added noise during the photon conversion process can be analyzed by calculating the power spectrum density (PSD) of the output fields. For a field $\hat{x}(t)$, the power spectrum density is given by $S_\mathrm{x}[\omega]=\frac{1}{2\pi}\int_{-\infty}^{+\infty}\mathrm{d}\omega'\langle\hat{x}^\dagger[-\omega]\hat{x}[\omega']\rangle$, where $\hat{x}[\omega]$ is the Fourier transform of $\hat{x}(t)$. The input thermal noise terms satisfy relations
\begin{equation}
\begin{split}
\langle\hat{b}_\mathrm{in2}[\omega]\hat{b}_\mathrm{in2}^\dagger[\omega']\rangle&=(\Bar{n}_\mathrm{th,m}+1)2\pi\delta(\omega+\omega'),  \\
\langle\hat{b}_\mathrm{in2}^\dagger[\omega]\hat{b}_\mathrm{in2}[\omega']\rangle&=\Bar{n}_\mathrm{th,m}2\pi\delta(\omega+\omega'),   \\
\langle\hat{c}_\mathrm{in2}[\omega]\hat{c}_\mathrm{in2}^\dagger[\omega']\rangle&=(\Bar{n}_\mathrm{th,e}+1)2\pi\delta(\omega+\omega'),  \\
\langle\hat{c}_\mathrm{in2}^\dagger[\omega]\hat{c}_\mathrm{in2}[\omega']\rangle&=\Bar{n}_\mathrm{th,e}2\pi\delta(\omega+\omega'),
\label{noise}
\end{split}
\end{equation}
with
\begin{equation}
\Bar{n}_\mathrm{th,m}=\frac{1}{e^{\hbar\omega_\mathrm{m}/k_\mathrm{B}T_\mathrm{m}}-1}, \quad
\Bar{n}_\mathrm{th,e}=\frac{1}{e^{\hbar\omega_\mathrm{e}/k_\mathrm{B}T_\mathrm{e}}-1}.
\end{equation}
Here, $T_\mathrm{m}$ and $T_\mathrm{e}$ are the temperatures of the intrinsic mechanical and microwave thermal baths, respectively. 
Note that we have neglected the thermal noise of the optical mode since it will be at the quantum ground state even at room temperature.

For \textbf{microwave-to-optical conversion (upconversion)}, the input fields are the microwave signal $\hat{c}_\mathrm{in1}$, the microwave noise $\hat{c}_\mathrm{in2}$, and the mechanical noise $\hat{b}_\mathrm{in2}$. Then the optical output PSD can be expressed as
\begin{equation}
\begin{split}
S_\mathrm{aout1}[\omega]&=\abs{\mathrm{S_{1,3}}}^2\frac{1}{2\pi}\int_{-\infty}^{+\infty}\mathrm{d}\omega'\langle\hat{c}_\mathrm{in1}^\dagger[-\omega]\hat{c}_\mathrm{in1}[\omega']\rangle  +  \abs{\mathrm{S_{1,4}}}^2\frac{1}{2\pi}\int_{-\infty}^{+\infty}\mathrm{d}\omega'\langle\hat{c}_\mathrm{in2}^\dagger[-\omega]\hat{c}_\mathrm{in2}[\omega']\rangle \\
&+\abs{\mathrm{S_{1,5}}}^2\frac{1}{2\pi}\int_{-\infty}^{+\infty}\mathrm{d}\omega'\langle\hat{b}_\mathrm{in2}^\dagger[-\omega]\hat{b}_\mathrm{in2}[\omega']\rangle.
\label{upcon}
\end{split}
\end{equation}
The first term is the converted optical signal, the second and the third terms are noise. The \textit{added noise} of the converter is defined as the noise referred to the input. Plugging in Eq.\,(\ref{noise}), we can get the microwave added noise $n_\mathrm{up,e}$ and the mechanical added noise $n_\mathrm{up,m}$ as
\begin{align}
n_\mathrm{up,e}&=\frac{\abs{\mathrm{S_{1,4}}}^2}{\abs{\mathrm{S_{1,3}}}^2}\Bar{n}_\mathrm{th,e}=\frac{\kappa_\mathrm{e,i}}{\kappa_\mathrm{e,c}}\Bar{n}_\mathrm{th,e},\\
n_\mathrm{up,m}&=\frac{\abs{\mathrm{S_{1,5}}}^2}{\abs{\mathrm{S_{1,3}}}^2}\Bar{n}_\mathrm{th,m}=\frac{\Big[\big(\frac{\kappa_\mathrm{e}}{2}\big)^2+(\omega-\omega_\mathrm{e})^2\Big]\kappa_\mathrm{m}}{g_\mathrm{em}^2\kappa_\mathrm{e,c}} \Bar{n}_\mathrm{th,m}.
\end{align}

In our experiment, the microwave thermal bath is estimated to have only a few photons ($\Bar{n}_\mathrm{th,e}\approx1.6$ at 1\,K, and 5.6 at 3\,K). Therefore $n_\mathrm{up,e}$ can be easily suppressed to below one when the microwave coupling port is very over-coupled ($\kappa_\mathrm{e,i}\ll\kappa_\mathrm{e,c}$). For the mechanical added noise, when $\omega=\omega_\mathrm{e}=\omega_\mathrm{m}$, we have $n_\mathrm{up,m}=\frac{1}{C_\mathrm{em}}\frac{\kappa_\mathrm{e}}{\kappa_\mathrm{e,c}}\Bar{n}_\mathrm{th,m}$. Namely, a large $C_\mathrm{em}$ and the over-coupling condition can suppress the mechanical added noise even if device is physically in a ``hot" thermal bath. These theoretical analyses, in fact, reveal the effect of radiative cooling---that is by over coupling the microwave mode to a cold bath, the electromechanical modes can be cooled to quantum ground state despite the large thermal occupancy of their physical environment.

For \textbf{optical-to-microwave conversion (downconversion)}, the input fields are the optical signal $\hat{a}_\mathrm{in1}$, the mechanical noise $\hat{b}_\mathrm{in2}$, and the microwave noises from both the intrinsic bath $\hat{c}_\mathrm{in2}$ and the coupling port $\hat{c}_\mathrm{in1}$. Hence, the microwave output PSD is
\begin{equation}
\begin{split}
S_\mathrm{cout1}[\omega]&=\abs{\mathrm{S_{3,1}}}^2\frac{1}{2\pi}\int_{-\infty}^{+\infty}\mathrm{d}\omega'\langle\hat{a}_\mathrm{in1}^\dagger[-\omega]\hat{a}_\mathrm{in1}[\omega']\rangle + \abs{\mathrm{S_{3,5}}}^2\frac{1}{2\pi}\int_{-\infty}^{+\infty}\mathrm{d}\omega'\langle\hat{b}_\mathrm{in2}^\dagger[-\omega]\hat{b}_\mathrm{in2}[\omega']\rangle  \\
&+ \abs{\mathrm{S_{3,4}}}^2\frac{1}{2\pi}\int_{-\infty}^{+\infty}\mathrm{d}\omega'\langle\hat{c}_\mathrm{in2}^\dagger[-\omega]\hat{c}_\mathrm{in2}[\omega']\rangle  
+ \abs{\mathrm{S_{3,3}}}^2\frac{1}{2\pi}\int_{-\infty}^{+\infty}\mathrm{d}\omega'\langle\hat{c}_\mathrm{in1}^\dagger[-\omega]\hat{c}_\mathrm{in1}[\omega']\rangle.
\label{dncon}
\end{split}
\end{equation}
The first term is the converted microwave signal, the second and the third terms are the mechanical and the microwave noises. The last term is the noise coming in from the microwave input port, which then gets reflected and comes out as part of the microwave output. For practical quantum operation, a circulator is needed to separate the input and the output fields of the microwave port. The microwave input should be thermalized at $\sim$\,mK environment to make sure this last term in Eq.\,(\ref{dncon}) contributes zero noise. This is also consistent with our idea and experimental implementation of the radiative cooling \cite{Xu_cool}. 
Similarly, the microwave and the mechanical added noise during the downconversion can be obtained as
\begin{align}
n_\mathrm{down,e} &= \frac{\abs{\mathrm{S_{3,4}}}^2}{\abs{\mathrm{S_{3,1}}}^2}\Bar{n}_\mathrm{th,e} = \frac{\abs{g_\mathrm{om}^2   +   \big[\frac{\kappa_\mathrm{o}}{2}-i(\omega+\Delta_\mathrm{o})\big]         \big[\frac{\kappa_\mathrm{m}}{2}-i(\omega-\omega_\mathrm{m})\big]}^2      \kappa_\mathrm{e,i}}{g_\mathrm{em}^2 g_\mathrm{om}^2 \kappa_\mathrm{o,c}} \Bar{n}_\mathrm{th,e},\\
n_\mathrm{down,m} &= \frac{\abs{\mathrm{S_{3,5}}}^2}{\abs{\mathrm{S_{3,1}}}^2}\Bar{n}_\mathrm{th,m} =       \frac{\Big[\big(\frac{\kappa_\mathrm{o}}{2}\big)^2+(\omega+\Delta_\mathrm{o})^2\Big]\kappa_\mathrm{m}}{g_\mathrm{om}^2\kappa_\mathrm{o,c}} \Bar{n}_\mathrm{th,m}.
\end{align}

We see that the thermal noises contribute differently in different conversion directions. When $\omega=\omega_\mathrm{e}=\omega_\mathrm{m}=-\Delta_\mathrm{o}$, the microwave added noise simplifies to $n_\mathrm{down,e} = \frac{(C_\mathrm{om}+1)^2}{C_\mathrm{om}C_\mathrm{em}} \frac{\kappa_\mathrm{o}}{\kappa_\mathrm{o,c}} \frac{\kappa_\mathrm{e,i}}{\kappa_\mathrm{e}}   \Bar{n}_\mathrm{th,e}$. 
As we have discussed in the main text, a high conversion efficiency requires large and matched $C_\mathrm{om}$ and $C_\mathrm{em}$, so the first factor would be around one. An over-coupled optical port will reduce the second factor to close to one. Therefore, it is again very crucial to over couple the microwave port ($\kappa_\mathrm{e,i}/\kappa_\mathrm{e}\ll1$) to suppress $\Bar{n}_\mathrm{th,e}$. 
The mechanical added noise at $\omega=\omega_\mathrm{m}=-\Delta_\mathrm{o}$ becomes $n_\mathrm{down,m}=\frac{1}{C_\mathrm{om}}\frac{\kappa_\mathrm{o}}{\kappa_\mathrm{o,c}}\Bar{n}_\mathrm{th,m}$. Hence, it is important to have a large $C_\mathrm{om}$ to suppress the mechanical noise during the optical-to-microwave photon conversion.

\end{document}